\documentclass[a4paper]{jpconf}
\usepackage{graphicx}
\usepackage{float}
\usepackage{color}
\usepackage{soul}
\usepackage{array}
\usepackage{placeins}

\newcommand{\beq}{\begin{equation}}
\newcommand{\eneq}{\end{equation}}
\newcommand{\be}{\begin{equation}}
\newcommand{\ee}{\end{equation}}
\newcommand{\bea}{\begin{eqnarray}}
\newcommand{\eea}{\end{eqnarray}}
\newcommand{\ket}{\rangle}
\newcommand{\bra}{\langle}

\usepackage{wasysym}

\begin{document}
\title{Out of equilibrium charge transport in molecular electronic devices}
\author{Andrea Nava, Domenico Giuliano, Luca Lepori and Marco Rossi}
\address{Dipartimento di Fisica, Universit\`a della Calabria, Arcavacata di Rende I-87036, Cosenza, Italy \\
INFN - Gruppo collegato di Cosenza,
Arcavacata di Rende I-87036, Cosenza, Italy}
\ead{andrea.nava@fis.unical.it}

\begin{abstract}
Using the Lindblad equation approach, we study the nonequilibrium stationary state of a benzene ring connected to two reservoirs in 
the large bias regime, a prototype of a generic molecular electronic device. We show the emergence of an optimal working point (corresponding to
 a change in the monotonicity of the stationary current, as a function of the applied bias) and its robustness against chemical potential and 
 bond disorder.
\end{abstract}

\section{Introduction}
\label{intro}

Molecular electronics is a promising theoretical and experimental research field focusing on the possibility to implement single molecules in quantum circuit in order to realize molecular switches, logic gate, and junctions. It is, therefore, of the utmost importance to provide models describing 
  molecules  attached to two, or more, macroscopic contacts
gates and the currents across as a function of the voltage bias or of the temperature gradient between the contacts.

 In this manuscript, we model   the charge transport mechanisms in such systems using the   master equation formalism, already 
 implemented for one-dimensional chains \cite{rossini_0, rossini, lindblad_nrg}. To be specific, we resort to the Lindblad equation (LE) approach to open quantum systems. This approach allows to derive the effective dynamics of the system by integrating 
over the reservoir degrees of freedom under the Markovian approximation,  consisting in 
 neglecting memory effects of the reservoirs. 
In the following, we first describe the general method and then apply it to a simple paradigmatic example, the benzene ring.

\section{The model}
\label{lind}

The starting point is the H\"uckel Hamiltonian for the $\pi$-electrons of a generic $N$-atom molecule,
\beq
H   =   -\sum_{j \neq k=1}^{N}t_{j,k}c_{j}^{\dagger}c_{k}+\sum_{j=0}^{N}\epsilon_{j}c_{j}^{\dagger}c_{j}
\:\:\:\: .
\label{eq:deltastar-H}
\eneq
\noindent
In Eq.[\ref{eq:deltastar-H}],   $c_j^\dagger , c_j $ are   single-electron creation and annihilation operators 
at molecular site $j$, satisfying  the canonical anticommutation relations $\{ c_j , c_{k}^\dagger \} = \delta_{j,k}$,
$t_{j,k}$ is the single-electron hopping strength between the $j$-th and the $k$-th atoms and $\epsilon_j$ is the on-site energy. For 
the sake of simplicity, we neglect the interaction between electrons, which can be easily implemented following Ref.\cite{lindblad_nrg}. 

We are interested in studying the behavior of the molecule when it is connected to two or more external reservoirs through some of its atoms. The 
 LE   consists of  a first order
differential equation for the time evolution of the system density matrix $\rho(t)$, 
given by 
\beq
\dot{\rho} ( t  )=
-i [H,\rho (t)  ]+\sum_{k} (L_{k}\rho ( t )  L_{k}^{\dagger}-\frac{1}{2} \{ L_{k}^{\dagger}L_{k},\rho (t)  \}  )
\:\:\:\: . 
\label{eq:lindbladeq}
\eneq
\noindent
The first term at the right-hand side of Eq.[\ref{eq:lindbladeq}]  is the
Liouvillian that describes the unitary evolution determined  by the Hamiltonian, $H$. The second term, the Lindbladian, includes
dissipation and decoherence on  the system dynamics introduced by the reservoirs. It 
depends on the so-called  ``jump'' operators $L_k$, which are determined by the
coupling between the system and the reservoirs.

In the following, we   consider  reservoirs that locally inject fermions to,  or extract fermions from, a generic site $j$ of 
the molecule, at  given and fixed rates. We describe the injecting and extracting reservoirs at site $j$ in 
terms of the   Lindblad operators $L_{in,j} $ and $L_{out,j} $, given by 
$ L_{in,j} = \sqrt{\Gamma_{j}}c_{j}^{\dagger}$, $L_{out,j} = \sqrt{\gamma_{j}}c_{j}$, with $\Gamma_j$  and $\gamma_j$  being  the coupling strengths respectively determining 
the creation and the    annihilation   of a fermion at  site $j$.
Once we  determine  $\rho ( t )$ by solving Eq.[\ref{eq:lindbladeq}], we compute 
the (time dependent) expectation value of any observable $O$ 
using $\bra O(t) \ket = {\rm Tr} [ O \rho\left(t\right) ] $.
For the sake of our analysis, we compute the currents flowing from the reservoirs into the site $j$,  $I_{in,j} ( t ) $,  or from site $j$ to 
the reservoir, $I_{out,j} ( t )$.  These are given by $I_{in,j} ( t )  = \Gamma_{i} (1-  \bra n_j ( t ) \ket)$ and $I_{out,j}( t ) = \gamma_{i}  \bra n_j ( t ) \ket$, with $n_j\left( t \right) = c^{\dagger}_j c_j$ the number operator. 
The net current exchanged at time $t$ 
between the reservoirs and the site $j$ is given by $I_{j} (t)  =  I_{in,j} ( t ) - I_{out,j} ( t )$.

Using Eq.[\ref{eq:lindbladeq}], we compute the current across a small Benzene ring connected to two reservoirs. The system asymptotically evolves to a non-equilibrium steady state (NESS), which we determine from the condition $\dot{\rho} = 0$. In such a state we compute the stationary current in order to find the optimal working point (OWP), i.e. the change in the monotonicity of the stationary 
current as a function of the applied bias, for a molecular electronic device.

\section{Benzene ring}
\label{benzene}

\begin{figure}[h]
\center
\includegraphics*[width=16pc]{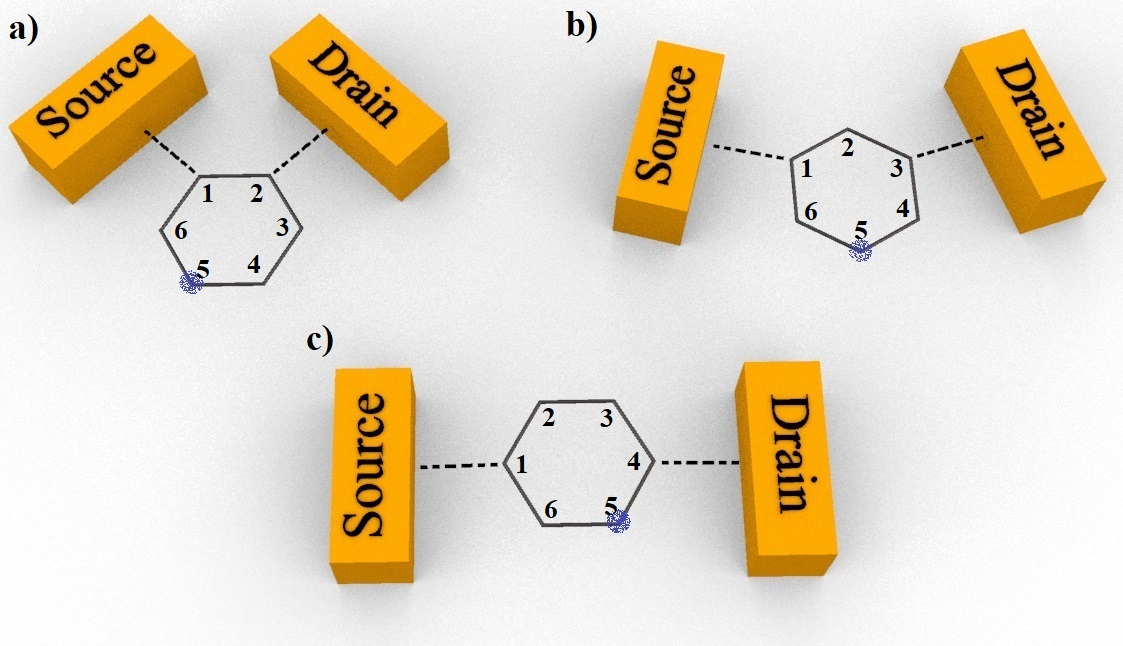}\hspace{2pc}%
\begin{minipage}[b]{16pc}\caption{\label{fig:benzene} A Benzene molecule connected to a source and drain reservoir in the a) ortho, b) meta and c) para configurations. Blue dot represents the gate voltage $\epsilon_g$.}
\end{minipage}
\end{figure}

The H\"uckel Hamiltonian for the Benzene ring is given by \cite{benzene_1}

\begin{equation}
H_{B}=\left(\begin{array}{cccccc}
\epsilon_{1} & -t_{12} & 0 & 0 & 0 & -t_{16}\\
-t_{21} & \epsilon_{2} & -t_{23} & 0 & 0 & 0\\
0 & -t_{32} & \epsilon_{3} & -t_{34} & 0 & 0\\
0 & 0 & -t_{43} & \epsilon_{4} & -t_{45} & 0\\
0 & 0 & 0 & -t_{54} & \epsilon_{5} & -t_{56}\\
-t_{61} & 0 & 0 & 0 & -t_{65} & \epsilon_{6} 
\end{array}\right) \:\:\: ,
\end{equation}
\noindent
whith real hopping amplitude, $t_{ij}=t^*_{j,i}$, between consecutive atoms of the ring. In the following we set $t_{ij}=t=2eV$ and $\epsilon_j=0$ except for $j=5$, where a local gate potential, $\epsilon_g$, is applied (values are taken from Ref.\cite{benzene_1}).
We connect the Benzene molecule to two reservoirs, one located on the site $1$ and another located on the site $L$ with $L=2$ for the ortho-configuration, $L=3$ for the meta-configuration and $L=4$ for the para-configuration (see Fig.\ref{fig:benzene} for a pictorial representation of the setup).

\noindent

\begin{figure}[h]
\begin{center}
\includegraphics*[width=1. \linewidth]{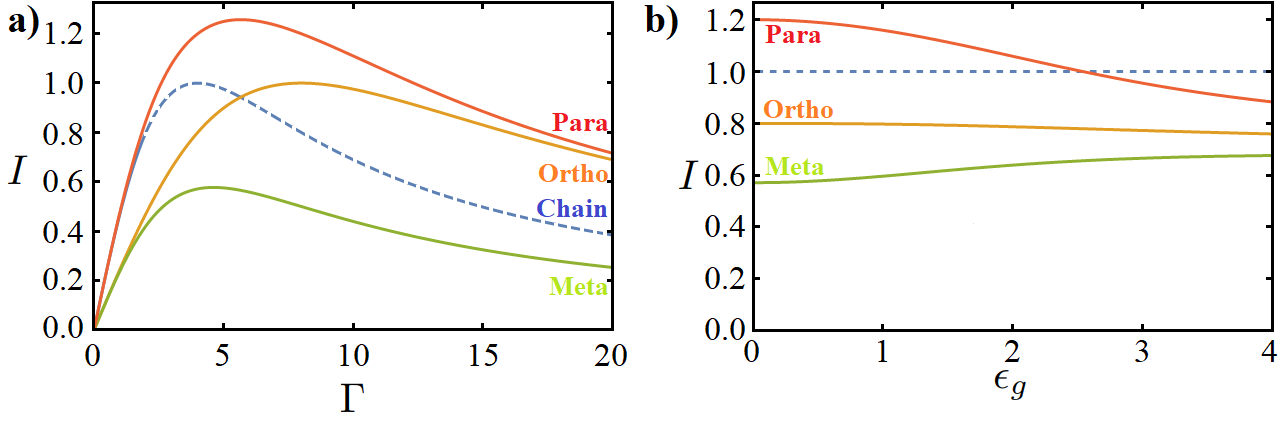}
\caption{\label{fig:large_bias_benzene} Panel a) current across a benzene molecule in terms of the coupling strength $\Gamma=\Gamma_1=\gamma_L$ in the large bias regime for the three configurations. Dashed blue curve corresponds to the current across a linear chain of $6$ sites. We set $t=2eV$ and $\epsilon_g=0$. The maximum of the curve is the OWP \cite{lindblad_nrg}. \\
Panel b) current across a benzene molecule in terms of the gate voltage $\epsilon_g$ in the large bias regime. Dashed blue curve corresponds to the current across a chain of $6$ sites (no gate voltage applied). We set $t=2eV$ and $\Gamma_1=\gamma_L=4$ corresponding to the OWP for the linear chain.}
\end{center}
\end{figure}

As a paradigmatic regime, we consider  
the large bias limit, in which
one of the reservoirs, the one coupled to the site $1$,  acts as an electron ``source'', 
by only injecting electrons in the system ($\gamma_1=0$), and the other, the one coupled to the site $L$, 
acts as an electron ``drain'',  by only  absorbing electrons from the system ($\Gamma_L=0$). As a result, 
electrons entering at the site 1 must travel all the way down to the site $L$, before leaving the system. Accordingly, the  boundary dynamics is determined only 
by the coupling strengths, $\Gamma_1=\gamma_L \equiv \Gamma$, while the bulk dynamics only depends on the hopping strengths $t$ and the gate voltage $\epsilon_g$.
In Fig.[\ref{fig:large_bias_benzene}a] we plot the stationary current across the benzene ring, $I=I_{in,1}=-I_{out,L}$, against the coupling strength to the two reservoirs for the para, ortho and meta configurations, comparing it with the case of the one-dimensional chain discussed in Ref.\cite{lindblad_nrg}. Two interesting features emerge: first, the OWP, already observed for a linear chain \cite{lindblad_nrg}, is present also in the benzene ring; second, the para and ortho configurations exhibit higher values of the current compared to the one-dimensional chain. This suggests that the introduction of molecular electronic devices within a quantum network can, in principle, improve its performance. In Fig.[\ref{fig:large_bias_benzene}b] we compute the current across the ring as a function of the gate voltage $\epsilon_g$, at fixed applied bias. Interestingly, the stationary current of the three possible configurations show opposite behaviors as a function of  $\epsilon_g$, decreasing in the para configuration, increasing in the meta configuration and keeping almost constant in the ortho configuration.

\noindent
\begin{figure}
\center
\includegraphics*[width=1. \linewidth]{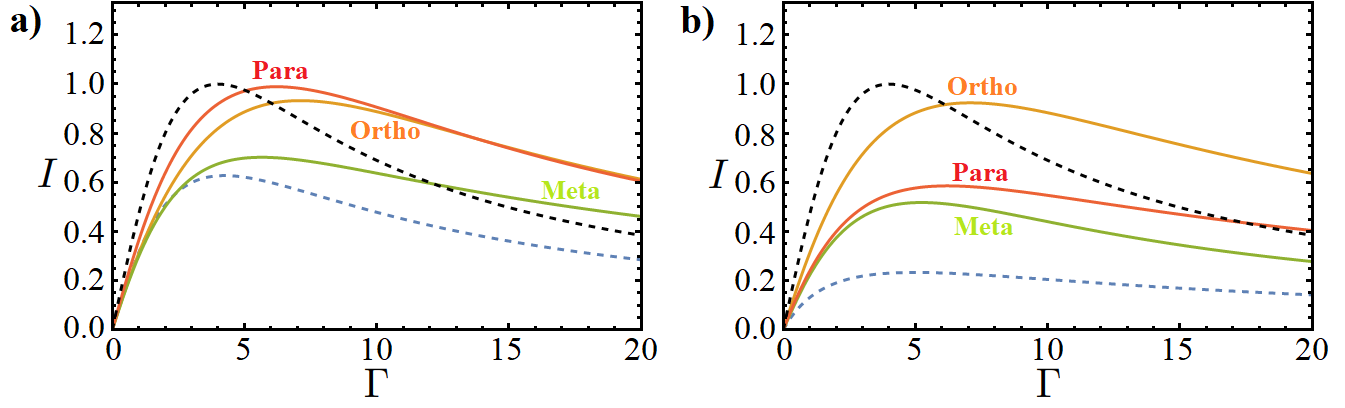}
\caption{\label{fig:lange_bias_benzene_disorder} Current across a disordered benzene molecule in terms of the coupling strength $\Gamma=\Gamma_1=\gamma_L$ in the large bias regime for the three configurations. Dashed black curve corresponds to the current across a clean chain of $6$ sites (like in Fig.[\ref{fig:large_bias_benzene}]), dashed blue curve to the current across a disordered chain of $6$ sites. We set $t=2eV$, $\epsilon_g=0$ and a) strong "chemical potential disorder", $\sigma_{\epsilon}=1.2$, $\sigma_t=0$, b) strong "bond disorder", $\sigma_{\epsilon}=0$, $\sigma_t=1.2$.}
\end{figure}
\noindent
An interesting test concerns the robustness of the OWP against disorder. To introduce disorder in the benzene ring we can, e.g., randomize the on-site energy $\epsilon$ and/or the bond electron hopping strength $t$. Disorder is, in general, expected to substantially affect the transport properties
of the system, especially in lower dimensions. In the following, we compare the current across the benzene ring, in the para, meta and ortho configurations, with the current across a one dimensional chain \cite{lindblad_nrg}. Technically, we realize this by 
setting  \cite{disorder_ring} $\epsilon_{\rm eff} \to \epsilon_{j}= \epsilon + \delta\epsilon_{j}$ and $t_{\rm eff} \to t_{i,j}= t + \delta t_{i,j}$
with $i<j=1,...,L$. The $\{ \delta \epsilon_j \}$ are independent 
random variables described by a probability  distribution $ P_{\epsilon} [ \{ \delta \epsilon_j \} ] 
= \prod_{ j = 1}^\ell p_{\epsilon} ( \delta \epsilon_j )$. Specifically, we choose $p_{\epsilon} ( \delta \epsilon )$ to be 
the  probability distribution for the $\{\delta \epsilon \}$, with average 
$\bar{\delta \epsilon} = \int \: d  x \: x p_{\epsilon} ( x ) = 0$, and with 
variance $\sigma_{\epsilon}^2 = \int \: d x \: x^2 p_{\epsilon} ( x )$.
We use the uniform probability distributions given by   
\beq
p_{\epsilon} ( \delta \epsilon ) \: = \:  \Biggl\{ \begin{array}{l}
\frac{1}{2 \sqrt{3} \sigma_{\epsilon}} \;\; , \; {\rm for} \: - \sqrt{3} \sigma_{\epsilon} \leq \delta \epsilon \leq \sqrt{3} \sigma_{\epsilon} \\
0 \:\: , \: {\rm otherwise}
                 \end{array}
\:\:\:\: , 
\label{disl.4}
\eneq

\noindent
with equivalent formulas holding for the hopping strength fluctuations {$\delta t_{i,j}$}. Having 
assumed the probability distribution in Eq.[\ref{disl.4}], we can 
estimate the disorder-averaged current distribution at given values of 
$\sigma_\epsilon$ and $\sigma_t$. 
In Fig.[\ref{fig:lange_bias_benzene_disorder}], we show the behavior of the stationary current, as a function of applied bias, in the presence of a strong chemical potential disorder (panel a) or a strong bond disorder (panel b). Interestingly, while the one-dimensional chain is strongly affected by disorder with a remarkable reduction of the current, the benzene ring is much more robust. Indeed, comparing panel a) and b), we observe that the para configuration is robust againts chemical potential disorder but is affected by the bond one, the meta configuration  current is enhanced by chemical disorder and the ortho configuration is quite unaffected by the disorder. These effects are expected to be a consequence of the interferences between the two branches of the ring.

\section{Conclusions}
\label{concl}

Using the LE framework, we have discussed the main features of the NESS arising in a small disordered molecular electronic device connected to two reservoirs in the large bias limit. The preliminary results shown in the manuscript suggest that the LE framework can be succesfully extended to non-linear systems, like molecular electronic devices, small DNA chains or junctions of quantum wires in the presence of two or more reservoirs in order to investigate their transport properties under the effect of a large voltage bias \cite{anulene, dna, ring, kondo_1, kondo_2}. Furthermore, the role of disorder and/or interaction can be discussed and, if a multi-site Lindblad bath is considered \cite{multi}, it is also possible to introduce a temperature bias to investigate the thermal transport across such devices and, for example, the breakdown of the Wiedemann-Franz law \cite{franz_1, franz_2}.

\ack
A. N. was financially supported  by POR Calabria FESR-FSE 2014/2020 - Linea B) Azione 10.5.12,  grant no.~A.5.1.
D. G., L. L. and M. R. acknowledge  financial support  from Italy's MIUR  PRIN projects TOP-SPIN (Grant No. PRIN 20177SL7HC).

\end{document}